\documentclass[aps,prd,preprint,showpacs,groupedaddress,showkeys]{revtex4}

\usepackage[colorlinks, linkcolor=blue, citecolor=blue, urlcolor=blue]{hyperref}
\usepackage{graphicx}
\usepackage{picins}
\usepackage{amsfonts}

\begin{document}

	\title{Black Hole--Never Forms, or Never Evaporates}

	\author{Yi Sun}
	\email{fantadox@gmail.com}
	\affiliation{Department of Modern Physics, University of Science and Technology of China, Hefei 230026, China}



	\begin{abstract}

Many discussion about the black hole conundrums, such as singularity and information loss, suggested that there must be some essential irreconcilable conflict between quantum theory and classical gravity theory, which cannot be solved with any semiclassical quantized model of gravity, the only feasible way must be some complete unified quantum theory of gravity.

In \cite{Vachaspati2007a}, the arguments indicate the possibility of an alternate outcome of gravitational collapse which avoids the information loss problem. In this paper, also with semiclassical analysis, it shows that so long as the mechanism of black hole evaporation satisfies a quite loose condition that the evaporation lifespan is finite for external observers, regardless of the detailed mechanism and process of evaporation, the conundrums above can be naturally avoided. This condition can be satisfied with Hawking-Unruh mechanism. Thus, the conflict between quantum theory and classical gravity theory may be not as serious as it seemed to be, the effectiveness of semiclassical methods might be underestimated.

An exact universal solution with spherical symmetry of Einstein field equation has been derived in this paper. All possible solutions with spherical symmetry of Einstein field equation are its special cases.

In addition, some problems of the Penrose diagram of an evaporating black hole first introduced by Hawking in 1975 \cite{Hawking1975} are clarified.

	\end{abstract}

	\maketitle

	\section{Introduction}

According to Oppenheimer \& Snyder\cite{Oppenheimer1939}, there are two influencing conclusions of black hole(BH) formation: ``the total time of collapse for an observer comoving with the stellar matter is finite... an external observer sees the star asymptotically shrinking to its gravitational radius.''

\begin{figure}[htb]
\centering
\includegraphics[scale=0.30, angle=0]{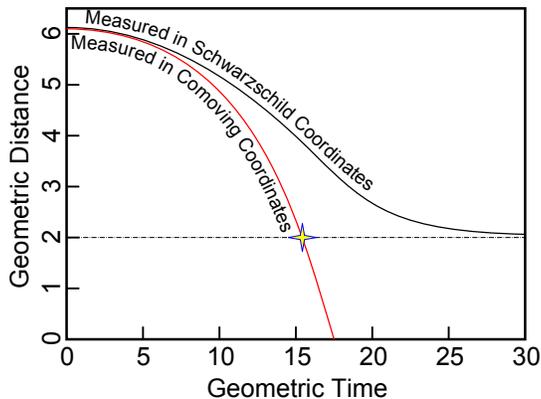}
\caption{Comparison of ingoing world line in different coordinates} \label{f:mislead}
\end{figure}

It must be mentioned that the word ``see'' means measuring by coordinates, not ``watching'' the light emitted from the star, which leads to a widespread misleading statement: what an external observer sees is just optical illusion, a BH has already formed, leaving only its light behind. It is difficult to trace back the origin of this misleading statement, but a figure in the early well-illustrated introduction of BH in \cite{Ruffini1971} may lead to some misunderstanding. One world line represented in two different coordinate systems of an infalling object are put together into one figure to show the difference between coordinate systems (see Fig.\ref{f:mislead}). However, it is easily misunderstood that the red curve is the real trajectory of the infalling object, and the black curve is the trajectory of its optical image.

\begin{figure}[htb]
\centering
\includegraphics[scale=0.35, angle=0]{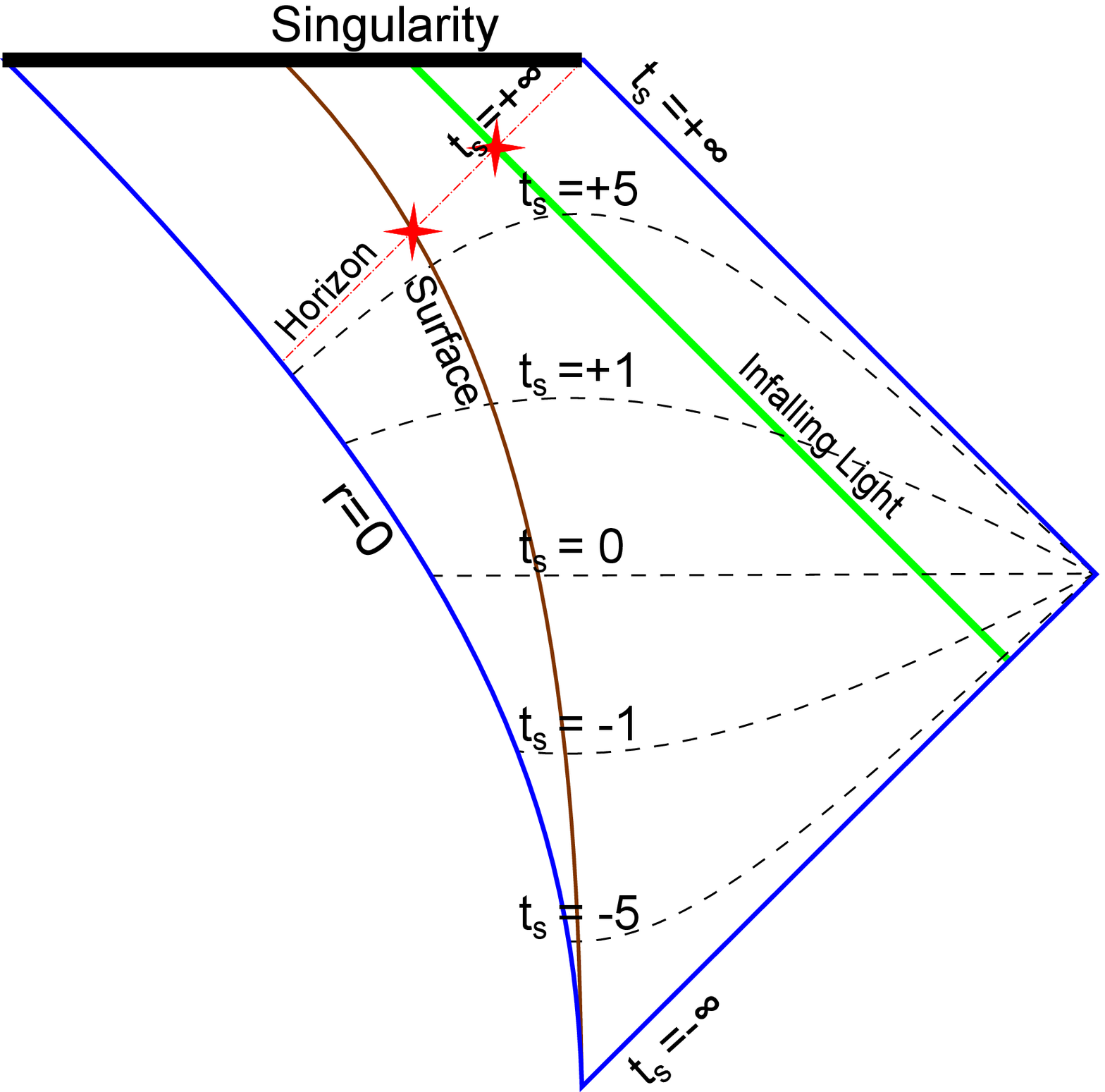}
\includegraphics[scale=0.30, angle=0]{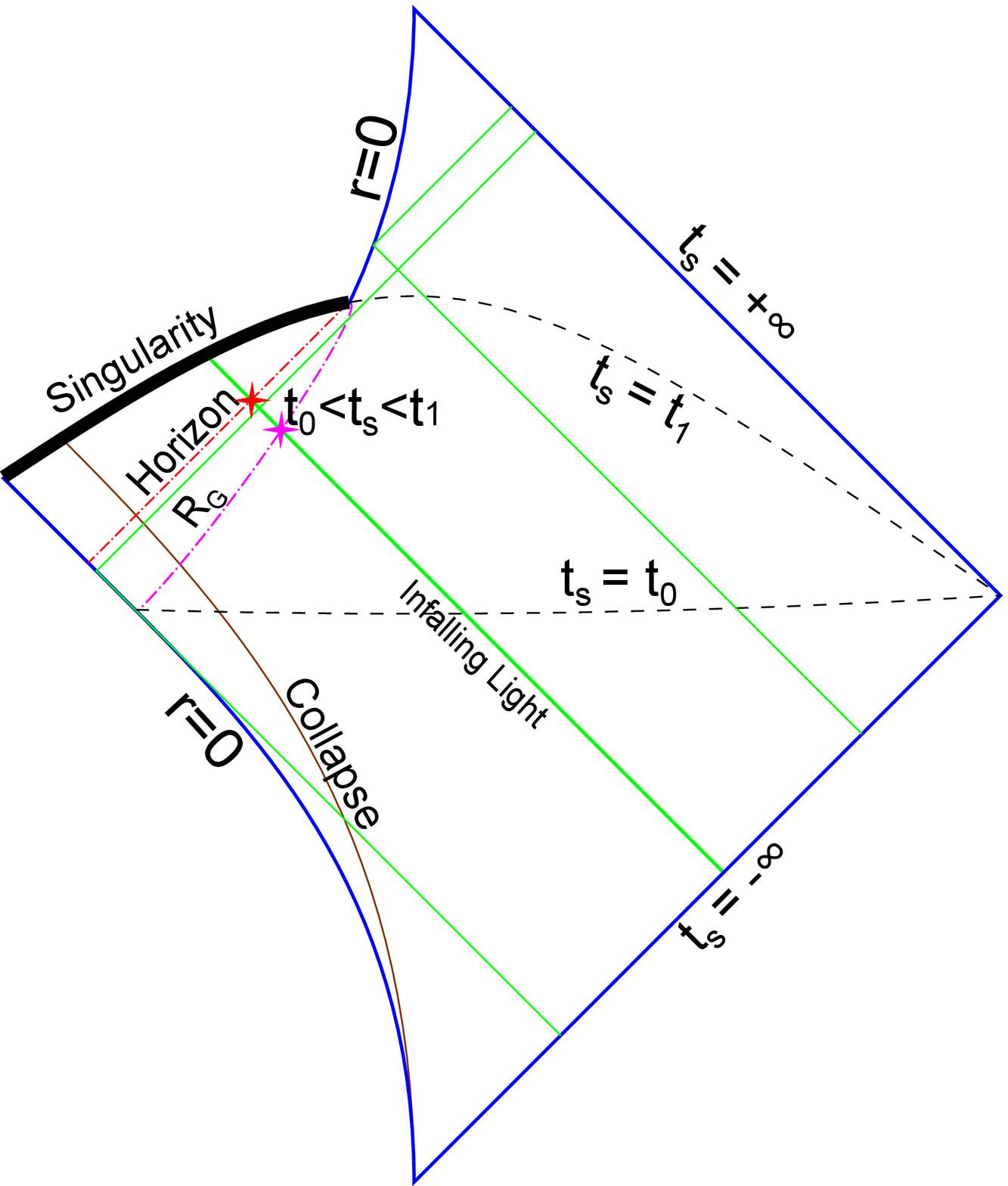}
\caption{Left: BH formation, Right: BH formation and evaporation} \label{f:BH}
\end{figure}

With Penrose diagram, the misleading statement can be easily clarified. Fig.\ref{f:BH} on left shows a collapsing star with the trajectory of its surface (brown) and an ingoing light (green), the black dashed lines represent curves of constant Schwarzschild time $t_S$. It is clearly that the Schwarzschild time $t_S$ of any event on horizon is infinite, but objects can still encounter the horizon within finite proper time, and the collapsing star can form within finite time in a comoving coordinates.\cite{Wald1994}

One may ask that who care whether the BH has really formed or not? A distant observer can never distinguish it from outside, and an infalling observer will encounter the horizon and the singularity in quite a short time. However, Vachaspati\cite{Vachaspati2007} pointed out that there would be significant difference for distant observers. The almost forming quasi-BH is called ``black star'' in his paper. It shows that in the case of two black stars collision, electromagnetic radiations will be emitted, which is sharply distinct from the perfect BHs that can only emit gravitational wave radiations.

Hawking's innovative paper \cite{Hawking1975}, with the semiclassical method of applying QFT in curved spacetime, shows that BHs will emit particles with a black-body spectrum that will cause BH evaporation. For a distant observer, a Schwarzschild BH with initial mass $m_0$ will have a temperature $T_H = (8\pi m_0)^{-1}$, and have a finite evaporation lifespan $t_{ev} = 5120\pi {m_0}^3$ ($G=c=\hbar=1$ units is adopted in this paper). This requires that the event that the BH vanishes must map to some finite Schwarzschild time, or no external observer could witness the event. For a typical BH with one solar mass ($m_0 \approx 2\times10^{30} kg$), the evaporation lifespan is about $10^{67}$ years, much longer than the current age of the universe about $10^{10}$ years. Furthermore, a BH will not emit but absorb radiation from environment until the background temperature becomes lower than its Hawking temperature $T_H$.

Hawking radiation caused a bunch of conundrums, such as the violation of baryon and lepton number conservation, the loss of quantum coherence etc.\cite{Wald1994}, referred to as information loss problem. Many people believes that these conundrums together with the singularity problem shows the essential irreconcilable conflict between quantum theory and classical gravity theory, which cannot be solved with any semiclassical quantized model of gravity, the only feasible way must be some complete unified quantum theory of gravity.

Fig.\ref{f:BH} on right is the refined version\cite{Lindesay2010} of the Penrose diagram which was first introduced in \cite{Hawking1975}, shows a BH with formation and evaporation process.

In \cite{Lindesay2008}, on BHs with finite evaporation lifespan, it states that ``the horizon of a static Schwarzschild geometry is a $t=\infty$ surface, while those of the dynamic BHs have surfaces of constant $t$ crossing the dynamic horizon.'' Some curves of constant Schwarzschild time (black dashed lines) are plotted in Fig.\ref{f:BH}.

This leads a problem that may be widely neglected: the surfaces of constant Schwarzschild time will cross the horizon of an evaporating BH. Which means all the world lines that encounter the horizon of a BH with finite evaporation lifespan, must also encounter the horizon at some finite Schwarzschild time, which is seriously violate the property of a classical eternal BH.

In addition, Hawking process shown in Fig.\ref{f:BH} on right has a bizarre behavior that the dominant energy condition has to be violated. There must be either negative or superluminal energy current between singularity and horizon because there is no time-like or null world line can transport energy from singularity to the exterior region of the BH. However, the energy of singularity has to be transported to the exterior region via Hawking radiation.

This paper, in Sec.\ref{ss:metric}, an exact universal solution with spherical symmetry of Einstein field equation(EFE) will be derived. In Sec.\ref{ss:lightcone} the solution will be applied to show that the light-cone forbids anything to encounter the horizon of an evaporating BH before it vanished. In Sec.\ref{ss:generalization}, a conjecture of general case will be proposed. In Sec.\ref{ss:conclusion} the conclusion will be illustrated with figures. Sec.\ref{s:summary} is a summary.

	\section{Semiclassical analysis of BH conundrums}

		\subsection{Exact universal solution with spherical symmetry} \label{ss:metric}

An exact universal solution with spherical symmetry of EFE will be derived in this section. Any physically permitted form of energy-momentum tensor with spherical symmetry has been involved in this solution, such as pulsing, collapsing, charge, radiation, etc. Any metric with spherical symmetry is a special case of this exact solution, such as Schwarzschild vacuum and interior metric, Reissner-Nordstrom(RN) metric.

Choosing the standard metric form under Schwarzschild coordinates\cite{Weinberg1972}:
\begin{displaymath}
ds^2 = -A(r,t)dt^2 + B(r,t)dr^2 + r^2 d\Omega^2
\end{displaymath}
where $d\Omega^2=d\theta^2 + \sin(\theta)^2 d\phi^2$.

The only non-vanishing elements of the Ricci tensor are:
\begin{displaymath}
\begin{array}{lll}
\mbox{Let~~} &\Delta &\equiv \frac{1}{4 A B}\left[2 A'' - A'\left(\frac{A'}{A} + \frac{B'}{B}\right) - 2\ddot{B} + \dot{B}\left(\frac{\dot{A}}{A} + \frac{\dot{B}}{B}\right)\right]\\
&R_{tt} &= A\left(\frac{1}{r B}\frac{A'}{A} + \Delta\right)\\
&R_{rr} &= B\left(\frac{1}{r B}\frac{B'}{B} - \Delta\right)\\
&R_{rt} &= \mbox{~}-\frac{\dot{B}}{r B}\\
&R_{\Omega\Omega} &= 1-\frac{1}{B}-\frac{r}{2 B} \left(\frac{A'}{A} - \frac{B'}{B}\right)
\end{array}
\end{displaymath}
where $R_{\Omega\Omega}=R_{\theta\theta}=R_{\phi\phi}/\sin(\theta)^2$. Convention: $\dot{X}\equiv\partial_t X$, $X'\equiv\partial_r X$.

Applying EFE: $R_{\mu\nu}-\frac{1}{2}R g_{\mu\nu} = 8\pi T_{\mu\nu}$, the only non-vanishing elements of the energy-momentum tensor are:
\begin{equation}
\begin{array}{rll}
A\rho &= T_{tt} &= \frac{1}{8\pi r^2}\left(r\frac{B'}{B}+B-1\right)\frac{A}{B}\\
B p &= T_{rr} &= \frac{1}{8\pi r^2}\left(r\frac{A'}{A}-B+1\right)\\
-\sqrt{A B} S &= T_{rt} &= \frac{1}{8\pi r}\frac{\dot{B}}{B}\\
r^2 p_t &= T_{\Omega\Omega} &= \frac{r^2}{32 \pi B}\left( \frac{\dot{B}}{A}\left(\frac{\dot{A}}{A}-\frac{\dot{B}}{B}\right)+2\frac{\dot{B}^2+B\ddot{B}}{A B} +\left(\frac{2}{r}-\frac{A'}{A}\right)\left(\frac{A'}{A}-\frac{B'}{B}\right)-2\frac{A'' B + A' B'}{A B}\right)
\end{array}
\label{eq:T}
\end{equation}
where $T_{\Omega\Omega}=T_{\theta\theta}=T_{\phi\phi}/\sin(\theta)^2$, $\rho(r,t)$ is the energy density, $S(r,t)$ is the radial energy flux density, $p(r,t)$ is the radial pressure, $p_t(r,t)$ is the tangential pressure.

The form of tangential pressure $p_t$ is quite complicated, but it is dependent with $\rho,p$, can be uniquely decided by $\rho,p$. It is noteworthy that it is not necessary to assume the pressure is isotropic that $p_t=p$.

From $T_{tt}$ and $T_{rr}$,
\begin{equation}
\begin{array}{l}
\displaystyle 8\pi r^2 B\rho = r\frac{B'}{B}+B-1\\
\displaystyle 8\pi r^2 B p = r\frac{A'}{A}-B+1
\end{array}
\label{eq:TttTrr}
\end{equation}

Adding up the two equations in Eqn.\ref{eq:TttTrr},
\begin{equation}
\begin{array}{l}
\displaystyle 8\pi r^2 B(\rho+p) = r\left(\frac{B'}{B}+\frac{A'}{A}\right)\\
\Rightarrow\displaystyle 8\pi r B(\rho+p) = \partial_r\ln(A B)\\
\Rightarrow\displaystyle A B = f_{\infty}(t)\exp{\left\{-\int\limits_{r}^{\infty} 8\pi \tilde{r} B(\tilde{r},t)\left(\rho(\tilde{r},t)+p(\tilde{r},t)\right) d\tilde{r}\right\}}
\end{array}
\label{eq:TttTrr_der}
\end{equation}
where $f_{\infty}(t)$ is an arbitrary function of $t$.

The solution with $A(\infty,t)=B(\infty,t)=1$ of Eqn.\ref{eq:TttTrr_der} is
\begin{equation}
\displaystyle A B = \exp{\left\{-\int\limits_{r}^{\infty} 8\pi \tilde{r} B(r,t)\left(\rho(\tilde{r},t)+p(\tilde{r},t)\right) d\tilde{r}\right\}}
\label{eq:AB}
\end{equation}

Let $f(r,t)\equiv\exp{\left\{-\int\limits_{r}^{\infty} 8\pi \tilde{r} B(r,t)\left(\rho(\tilde{r},t)+p(\tilde{r},t)\right) d\tilde{r}\right\}}$, thus $A(r,t)=f(r,t)/B(r,t)$.

Introducing the total energy function $M(r,t) = \int\limits_{0}^{r} 4\pi \tilde{r}^2 \rho(\tilde{r},t) d\tilde{r}$, thus
\begin{equation}
\begin{array}{l}
\displaystyle \rho(r,t) = \frac{M'}{4\pi r^2}\\
\displaystyle S(r,t) = -\sqrt{\frac{B}{A}}\frac{\dot{M}}{4\pi r^2}
\end{array}
\end{equation}

By the way, the radial velocity of an observer who sees $d M = M' d r + \dot{M} d t = 0$ can be defined as the radial velocity of the radial energy flux. For such an observer, the total energy inside his radial location is unchanging.
\begin{equation}
\begin{array}{l}
\displaystyle d M = M' d r + \dot{M} d t = 0\\
\Rightarrow\displaystyle \frac{d r}{d t} = -\frac{\dot{M}}{M'} \equiv v_r(r,t)\\
\Rightarrow\displaystyle \sqrt{\frac{B}{A}}v_r(r,t) = \frac{S}{\rho} \equiv u_r(r,t)
\end{array}
\label{eq:Ur}
\end{equation}
where $v_r$ is coordinate radial velocity of the energy flux and $u_r$ is measured by an observer rest at $r$.

From $T_{rt}$,
\begin{equation}
\begin{array}{l}
\displaystyle T_{rt} = -\sqrt{A B} S = B\frac{\dot{M}}{4\pi r^2} = \frac{1}{8\pi r}\frac{\dot{B}}{B}\\
\Rightarrow\displaystyle \frac{2\dot{M}}{r} = -\partial_t B^{-1}\\
\Rightarrow\displaystyle B(r,t)^{-1} = B_0(r)^{-1}-\int\limits_{0}^{t}\frac{2\dot{M}(r,\tilde{t})}{r}d\tilde{t} = B_0(r)^{-1}-\left(\frac{2M(r,t)}{r}-\frac{2M_0(r)}{r}\right)
\end{array}
\label{eq:Trt}
\end{equation}
where $M_0(r),B_0(r)$ are arbitrary functions of $r$.

The solution with $\lim\limits_{r\rightarrow\infty} B(r,t)^{-1} = 1-\frac{2M}{r}$ of Eqn.\ref{eq:Trt} is
\begin{equation}
B(r,t)^{-1}=1-\frac{2M(r,t)}{r}
\label{eq:B}
\end{equation}

Now, combining Eqn.\ref{eq:AB} and Eqn.\ref{eq:B}, the solution is complete:
\begin{equation}
\begin{array}{l}
\displaystyle ds^2 = -f(r,t)\left(1-\frac{2 M(r,t)}{r}\right)dt^2 + \left(1-\frac{2 M(r,t)}{r}\right)^{-1}dr^2 + r^2 d\Omega^2\\
f(r,t)=\exp{\left\{-\int\limits_{r}^{\infty} 8\pi \tilde{r}\left(1-\frac{2 M(\tilde{r},t)}{\tilde{r}}\right)^{-1}\left(\rho(\tilde{r},t)+p(\tilde{r},t)\right) d\tilde{r}\right\}}\\
M(r,t) = \int\limits_{0}^{r} 4\pi \tilde{r}^2 \rho(\tilde{r},t) d\tilde{r}
\end{array}
\label{eq:metric_nonvac}
\end{equation}

It is easily to validate the metric form by substituting into Eqn.\ref{eq:T}, $\rho, p$ are back, $p_t$ and $S$ are represented with $\rho,p$.

In this metric, the coordinate singularity is located at $r = 2 M(r,t)$, may be at not only one radius. Define $R_G(t)$:
\begin{equation}
R_G(t): \max r \in \{r = 2 M(r,t)\}
\label{def:R_G}
\end{equation}
there must be a BH if $R_G(t)>0$.

The total energy function $M(r,t)$ and the radial pressure distribution function $p(r,t)$ can be arbitrarily chosen under energy conservation law and some energy condition:
\begin{equation}
\begin{array}{ll}
M(0,t) = 0, M(\infty,t) = \mathcal{M}    & \mbox{(energy conservation law)}\\
\rho \ge 0, M(r,t) \ge 0                 & \mbox{(weak or dominant energy condition)}\\
\left|\frac{S}{\rho}\right|\le 1         & \mbox{(dominant energy condition)}\\
\rho + p \ge 0                           & \mbox{(null, weak, dominant or strong energy condition)}
\end{array}
\label{eq:cond_nonvac}
\end{equation}
Following the last condition, immediately:
\begin{equation}
0<f(r,t) \le 1 \mbox{~~for~~} r \ge R_G(t)
\label{eq:cond_f}
\end{equation}

In this paper, the exact form of $\rho(r,t),p(r,t)$ is irrelevant with the conclusion. In the sense of energy conditions, all physically permitted form of $\rho(r,t),p(r,t)$ will leads the same conclusion.

It is obviously that this metric will become the classical vacuum Schwarzschild metric under vacuum condition, and the Birkhoff's theorem will be satisfied automatically. As a further validation check of this general metric form of spherical symmetric field, the RN metric can be easily obtained by replacing $\rho$ and $p$ with the energy density and radial pressure of the static field produced by a point charge $\mathcal{Q}$, $\rho(r)=\frac{\mathcal{Q}^2}{8\pi r^4}$ and $p(r)=-\frac{\mathcal{Q}^2}{8\pi r^4}$:
\begin{equation}
\begin{array}{l}
f(r)=\exp{\left\{-\int\limits_{r}^{\infty} 8\pi \tilde{r}\left(1-\frac{2 M(\tilde{r})}{\tilde{r}}\right)^{-1}\left(\rho(\tilde{r})+p(\tilde{r})\right) d\tilde{r}\right\}}=1\\
M(r)=\int\limits_{0}^{r} 4\pi \tilde{r}^2 \rho(\tilde{r}) d\tilde{r}=\int\limits_{0}^{\infty} 4\pi \tilde{r}^2 \rho(\tilde{r}) d\tilde{r}-\int\limits_{r}^{\infty} 4\pi \tilde{r}^2 \rho(\tilde{r}) d\tilde{r}=\mathcal{M}-\frac{\mathcal{Q}^2}{2r}\\
\Rightarrow\displaystyle A(r)=B(r)^{-1}=1-\frac{2M(r)}{r}=1-\frac{2\mathcal{M}}{r}+\frac{\mathcal{Q}^2}{r^2}
\end{array}
\label{eq:metric_RN}
\end{equation}

	\subsection{The forbidden region of light cone} \label{ss:lightcone}

For a BH with finite evaporation lifespan, the only way to get in the horizon is getting in the horizon before it vanished, or it is impossible to encounter any horizon because there will be no more BH. This is quite different from the case of an eternal BH, in this case, even the horizon cannot be encountered at any finite coordinate time, but because the BH will never vanish, ingoing world lines can still encounter the horizon at infinite coordinate time, and in the comoving coordinates, an infalling observer will encounter the horizon within finite proper time.

In this section, the only assumption is that there has already been a Schwarzschild BH at the origin of Schwarzschild coordinates, and it will vanish after some finite Schwarzschild time $t_{ev}$, then there will be no more BH and horizon:
\begin{equation}
\begin{array}{l}
R_G(t) > 0 \mbox{~~~~~~~~~~if~~~~} t < t_{ev}\\
R_G(t) = 0 \mbox{~~~~~~~~~~if~~~~} t \ge t_{ev}
\end{array}
\label{eq:cond_evap}
\end{equation}
where the definition of $R_G(t)$ see Def.\ref{def:R_G}.

It is noteworthy that the contribution of radial radiation and radial matter flow has been naturally involved. The only requirement is the BH must vanish within finite time for external observer, and after that, there will be no more BH. The calculation is almost classical, and the only non-classical factor is the assumption of finite BH evaporation lifespan, which is the key to obtain the conclusion.

In addition, geodesics in non-vacuum region may be far different from the real trajectory of the particles, unless the particles do not participate with any other interactions but gravity. Considering the light cone, if the ingoing null geodesic cannot encounter the horizon, neither do anything.

For an ingoing null geodesic, radius $r$ is a function of Schwarzschild time $t$, $r=r(t), r(0) > R_G(0)$, thus $f(r,t)$ and $M(r,t)$ becomes function of $t$:
\begin{equation}
f(t) = f(r(t),t),~~~~ M(t) = M(r(t),t)
\label{eq:def_Mt}
\end{equation}
where $M(t)$ is just the total energy inside the current radius $r(t)$ of geodesic. It is noteworthy that $M(t)$ and $f(t)$ include not only the contribution of the energy inside the BH, but also the energy outside the BH, where the radiation of the BH has been naturally involved.

From metric Eqn.\ref{eq:metric_nonvac} the ingoing null geodesic becomes:
\begin{displaymath}
f(t)\left(1-\frac{2 M(t)}{r(t)}\right)dt^2 = \left(1-\frac{2 M(r)}{r(t)}\right)^{-1}dr^2
\end{displaymath}
for ingoing null geodesic, $dr/dt < 0$
\begin{equation}
\frac{dr}{dt} = - {f(t)}^{1/2}\left(1-\frac{2 M(t)}{r(t)}\right)
\label{eq:geo}
\end{equation}

The radial energy flux cannot travel faster than light, from Eqn.\ref{eq:Ur}:
\begin{equation}
\begin{array}{l}
\displaystyle\left|v_r(r,t)\right| \le \left|\frac{d r(t)}{d t}\right|\\
\displaystyle\Rightarrow\frac{\partial_t M(r,t)}{\partial_r M(r,t)} \le -\frac{dr(t)}{dt}\\
\displaystyle\Rightarrow\partial_t M(r,t)+{\partial_r M(r,t)} \frac{dr(t)}{dt} = \frac{d M(r(t),t)}{dt} = M'(t) \le 0
\end{array}
\label{eq:cond_M}
\end{equation}
so for ingoing null geodesic, the total energy inside radius $r(t)$ will never increase with $t$, because nothing can run out of the light cone.

Introducing reduced radius $\hat{r}$ and time $\hat{t}$:
\begin{equation}
\displaystyle \hat{r} = \frac{r(t)}{2 M(t)}, \mbox{~~~~~~~~} \hat{t} = \frac{t}{2 M(t)}
\end{equation}
for any $0 \le t < t_{ev}$, with Cond.\ref{eq:cond_evap},
\begin{equation}
\frac{d\hat{t}}{d t} = \frac{M(t) - M'(t) t}{2{M(t)}^2} > 0
\end{equation}
therefore $\hat{t}$ is a strictly monotone increasing function of any $0 \le t < t_{ev}$, the inverse function of $\hat{t}(t)$ exists.

Before the BH vanished, when there must be $M(t) > 0$ (see Def.\ref{eq:def_Mt}.), any finite Schwarzschild time $t$ maps to $\hat{t} = t/(2 M(t))$ also be finite, this means if any world line can encounter the horizon at any Schwarzschild time $t<t_{ev}$, the corresponding reduced time $\hat{t}$ must also be finite. When $t \ge t_{ev}$, there will be no more BH and horizon inside current radius $r(t)$.

Thus, regardless of the choice of coordinates, the necessary condition that a geodesic can encounter the horizon is its reduced radius $\hat{r}$ can reach $1$ within its finite reduced time $\hat{t} < \infty$.

According to Cond.\ref{eq:cond_M}, $M(\hat{t}) = M(t(\hat{t}))$ is a monotone non-increasing function of $\hat{t}$:
\begin{equation}
M'(\hat{t}) = \frac{d M(\hat{t})}{d\hat{t}} = M'(t)\frac{d t}{d\hat{t}} \le 0
\label{eq:cond_Mp}
\end{equation}
now Eqn.\ref{eq:geo} can be rewritten with $\hat{r}$ and $\hat{t}$:
\begin{equation}
\displaystyle
\frac{d\hat{r}}{d\hat{t}} = -{f(\hat{t})}^{1/2}\left(1-\frac{1}{\hat{r}}\right) -\left[{f(\hat{t})}^{1/2}\left(1-\frac{1}{\hat{r}}\right)\hat{t} + \hat{r}\right]\frac{M'(\hat{t})}{M(\hat{t})}
\label{eq:geo_r}
\end{equation}
the first term is the same as classical vacuum Schwarzschild metric, thus, the sign of the second term is the key. Before the BH vanished, if $r \ge 2 M(t)$, $\hat{r} \ge 1$, $\hat{t} \ge 0$, $f(\hat{t}) \ge 0$(Cond.\ref{eq:cond_f}), $M(\hat{t}) > 0$(Cond.\ref{eq:cond_nonvac}), $M'(\hat{t}) \le 0$(Cond.\ref{eq:cond_Mp}), then
\begin{displaymath}
-\left[{f(\hat{t})}^{1/2}\left(1-\frac{1}{\hat{r}}\right)\hat{t} + \hat{r}\right]\frac{M'(\hat{t})}{M(\hat{t})} > 0 \mbox{~~~~for~} \hat{r} > 1
\end{displaymath}

To decide whether the geodesic of Eqn.\ref{eq:geo_r} will encounter the horizon at some finite Schwarzschild time $t$, comparing Eqn.\ref{eq:geo_r} with the ingoing null geodesic in classical Schwarzschild metric, where $f(t)=1$, $M'(t) = M'(\hat{t}) = 0$:
\begin{equation}
\frac{d\hat{r}}{d\hat{t}} = -\left(1-\frac{1}{\hat{r}}\right)
\label{eq:geo_vac_r}
\end{equation}
this geodesic in classical vacuum Schwarzschild metric will never encounter the horizon at any finite $\hat{t}$, and so as Schwarzschild time $t$. It is noteworthy that for a classical eternal BH, this does not mean that the geodesic will not encounter the horizon. However, this is irrelevant here because there will be no more BH to encounter after $t_{ev}$. Now compare Eqn.\ref{eq:geo_r} and Eqn.\ref{eq:geo_vac_r}:
\begin{equation}
\begin{array}{lcl}
\displaystyle \frac{d\hat{r}}{d\hat{t}} & = & \displaystyle -{f(\hat{t})}^{1/2}\left(1-\frac{1}{\hat{r}}\right) -\left[{f(\hat{t})}^{1/2}\left(1-\frac{1}{\hat{r}}\right)\hat{t} + \hat{r}\right]\frac{M'(\hat{t})}{M(\hat{t})} \\
\displaystyle & \ge & \displaystyle -{f(\hat{t})}^{1/2}\left(1-\frac{1}{\hat{r}}\right) \\
\displaystyle & \ge & \displaystyle -\left(1-\frac{1}{\hat{r}}\right)
\end{array}
\end{equation}
this result shows that the geodesic of Eqn.\ref{eq:geo_r} will also never encounter the horizon at any finite $\hat{t}$, because the decreasing speed of $\hat{r}$ with $\hat{t}$ is no faster than the case of classical vacuum Schwarzschild metric.

Thus, the geodesic of Eqn.\ref{eq:geo_r} will never encounter the horizon at all, regardless of the choice of coordinates. Because if the geodesic cannot encounter the horizon before the BH vanished, then there will be no more BH and horizon to encounter.

To sum up, for an evaporating spherical symmetric BH with finite evaporation lifespan, the light-cone forbids anything to encounter the horizon, regardless of the choice of coordinates, even regardless of how the evaporation process contributes to the spacetime geometry, such as the contributions of the back-reaction of radiations and any other radial energy flux.

What would an evaporating BH look like as seen by an infalling observer? Comparing with the conclusion made with classical general relativity(GR) in \cite{Oppenheimer1939}, the total time of evaporation for an observer infalling to the BH is too short to enter the horizon, while an external observer sees the BH having an extremely long evaporation lifespan. As the infalling observer getting close to the horizon, he will see the horizon becomes incredible bright and shrinking faster and faster. With a terrible explosion, the BH vanishes before he enters its horizon. More information about Hawking radiation as seen by an infalling observer can be found in \cite{Greenwood2009}, but it is noteworthy that the effect of shrinking horizon has not been included.

		\subsection{Generalization} \label{ss:generalization}

\parpic{
\includegraphics[scale=0.30]{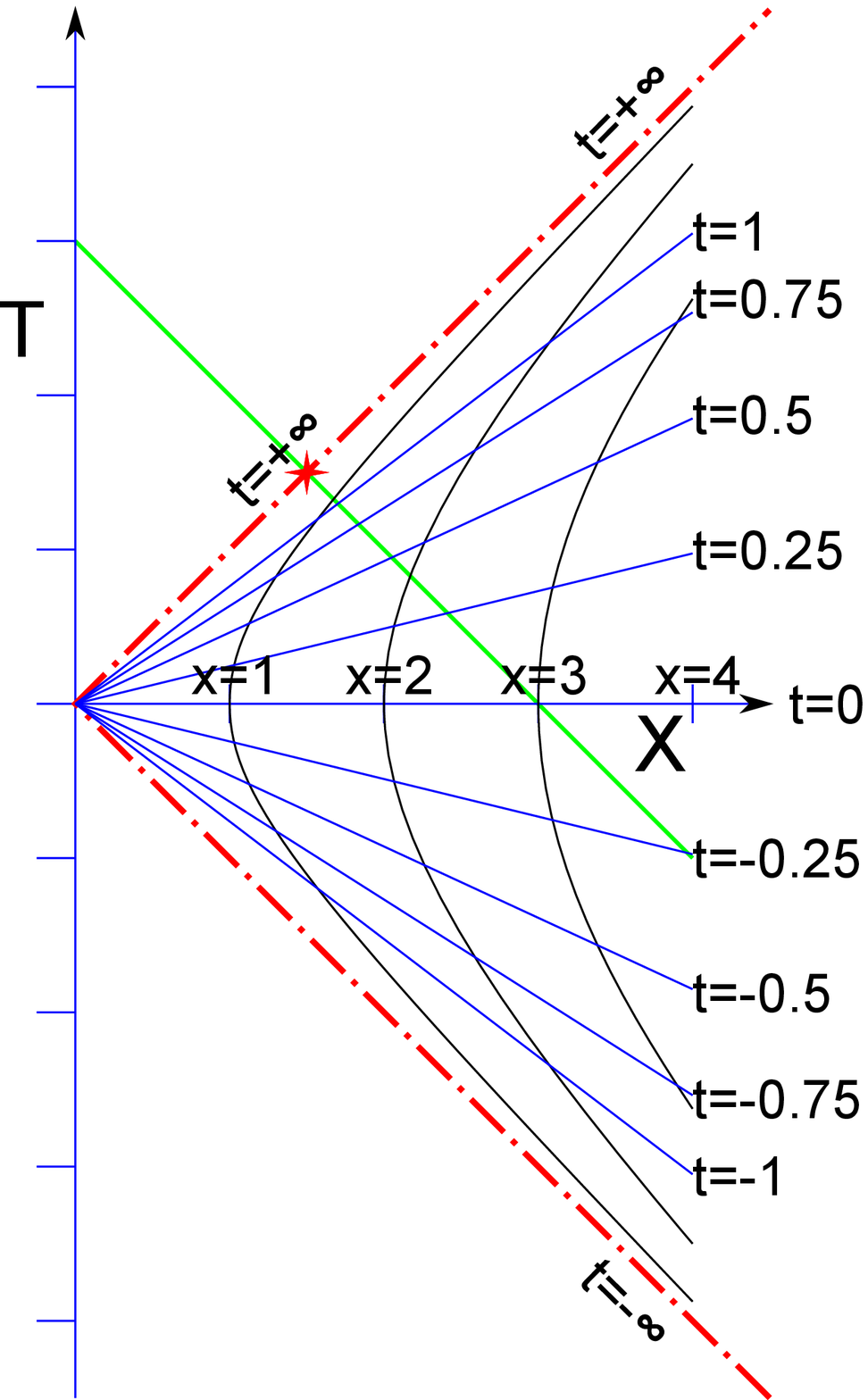}
}
The calculation in Sec.\ref{ss:lightcone} is valid for any spherical symmetric BH with finite evaporation lifespan. Nevertheless, what about the other general BHs? It is not so smart to solve special BHs one by one.

Heuristically, the result obtained in Sec.\ref{ss:lightcone} could be acquired more generally with Rindler geometry near the horizon.

If the near-horizon geometry is Rindler geometry, a hovering observer close to horizon is a Rindler observer. For a Rindler observer, nothing can encounter the horizon within finite Rindler time. The Rindler time on the horizon must also be infinite. However, the Penrose diagram of a BH with finite evaporation lifespan shown in Fig.\ref{f:BH} on right requires that the event on horizon must map to some finite Rindler time before the BH vanished, which strongly unfavor Fig.\ref{f:BH} on right.

The rigorous proof for the general cases may have to involve some global differential geometry technique. Nevertheless, along the heuristic thread above, it is reasonable to conjecture that for any BH with finite evaporation lifespan, nothing is physically permitted to encounter the horizon, regardless of the choice of coordinates, even regardless of how the evaporation process contributes to the spacetime geometry.

		\subsection{Causal structure and conclusion} \label{ss:conclusion}

\begin{figure}[htb]
\centering
\includegraphics[scale=0.50, angle=0]{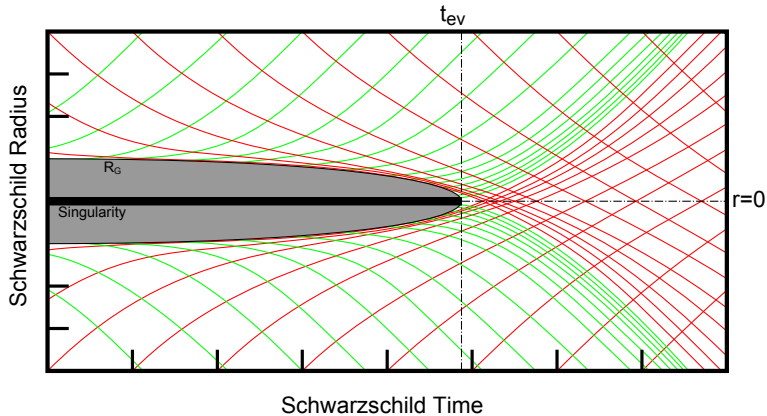}
\caption{Null geodesics of an evaporating BH in Schwarzschild coordinates} \label{f:BH_evap_Schw}
\end{figure}

Fig.\ref{f:BH_evap_Schw} is the plot of the null geodesics of an evaporating spherical BH in Schwarzschild coordinates. The solid red curves are the ingoing null geodesics (and eventually escaped from coordinates origin after the BH vanished), solid green curves are the radiations emitted near the horizon.

Fig.\ref{f:BH_evap_Schw} shows no ingoing (eventually escaping) null geodesics will terminate at singularity; therefore, these geodesics are complete, because the ingoing null geodesics cannot encounter the horizon within finite Schwarzschild time $t_{ev}$. Nevertheless the geodesics of the radiations emitted from horizon are not complete, the BH is pre-existing as assumed, which means the outgoing geodesics of radiations will encounter the coordinates singularity at past infinity.

\begin{figure}[htb]
\centering
\includegraphics[scale=0.50, angle=0]{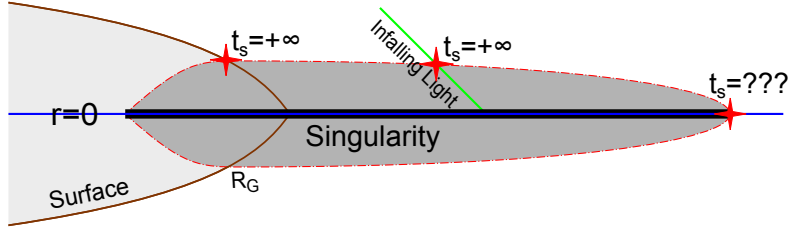}
\caption{Ingoing Finkelstein diagram of an evaporating BH (incorrect)} \label{f:BH_evap_Fink_Wr}
\end{figure}

One may ask what about ingoing Eddington-Finkelstein coordinates. It can be found in Birrell \& Davis\cite{Birrell1982}, shown in Fig.\ref{f:BH_evap_Fink_Wr}.

\begin{figure}[htb]
\centering
\includegraphics[scale=0.70, angle=0]{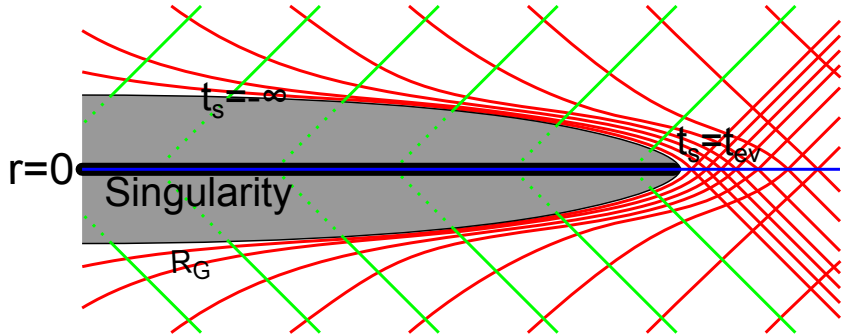}
\caption{Outgoing Finkelstein diagram of a pre-existing evaporating BH} \label{f:BH_evap_Fink}
\end{figure}

Fig.\ref{f:BH_evap_Fink_Wr} is exactly equivalent to the Penrose diagram in Fig.\ref{f:BH} on right, so they share exactly the same problem. Then what is the correct Finkelstein diagram of a pre-existing evaporating BH? The answer is it is impossible to plot a pre-existing evaporating BH in the ingoing Finkelstein diagram, but it is possible to plot in the outgoing Finkelstein diagram, which is used to describe WHs. However, in outgoing Finkelstein diagram, a WH cannot have a beginning, must be pre-existing. As shown in Fig.\ref{f:BH_evap_Fink}, the solid green curves are radiation emitted from the horizon; the solid red curves are ingoing lights.

Similarly, just like BH evaporation process cannot be shown in the ingoing Finkelstein diagram, the WH formation process also cannot be shown in the outgoing Finkelstein diagram, and, it is impossible to show both the formation and the evaporation process of a BH or a WH in one Finkelstein diagram. Although it is possible to show a pre-existing evaporating BH as a WH in outgoing Finkelstein diagram, there is still a problem that how a BH with a finite evaporation lifespan have existed for infinite long time (measured by external observer)?

In \cite{Hawking1976}, section IV.``WHITE HOLES'', Hawking considered a gedanken experiment about the arrow of time, it shows that there is no difference between BHs and WHs by means of ergodic assumption. As the time-reverse of an evaporating BH, a WH can also absorb radiation continuously from the environment, even can be created from gravitational collapse of interstellar radiations.

\begin{figure}[htb]
\centering
\includegraphics[scale=0.35, angle=0]{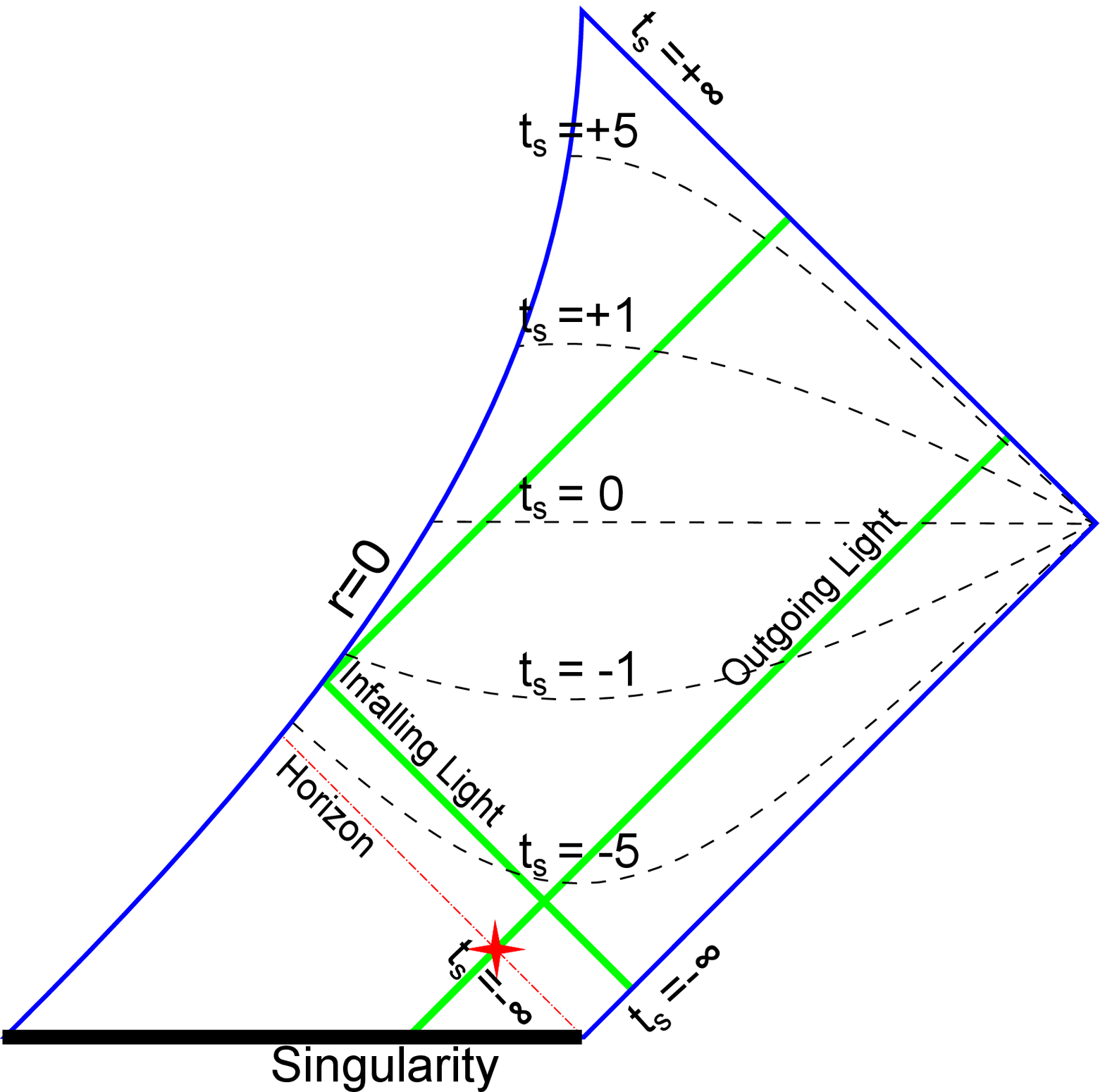}
\includegraphics[scale=0.30, angle=0]{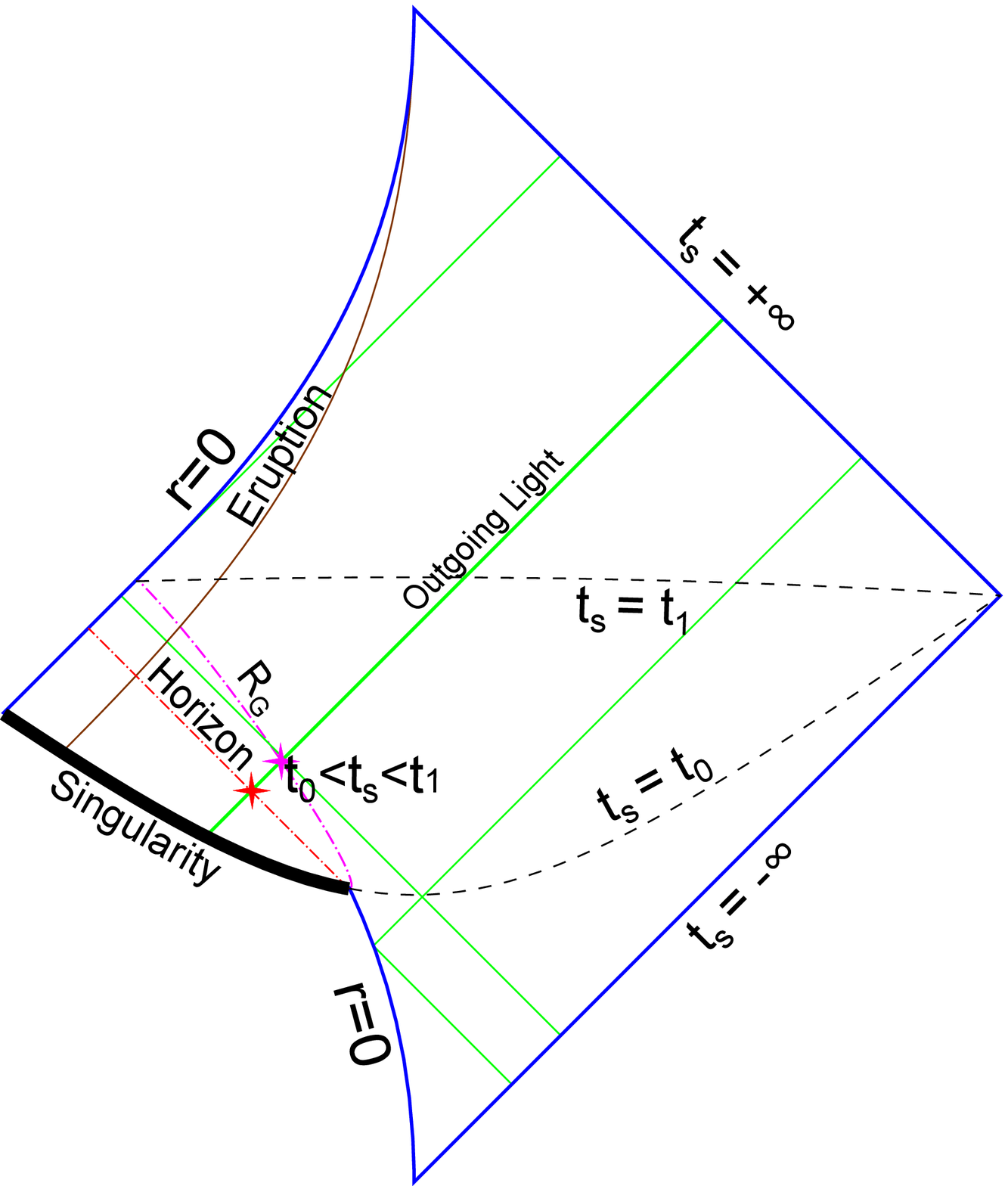}
\caption{Left: Pre-existing evaporating BH as an erupting WH, Right: WH with a beginning} \label{f:WH}
\end{figure}

Along the thread above, the Penrose diagram of a pre-existing BH with finite evaporation lifespan must be like an erupting WH, as shown in Fig.\ref{f:WH} on left, the exactly up-side-down version of Fig.\ref{f:BH} on left.

The Penrose diagram of an erupting WH with a beginning corresponding to the up-side-down version of Fig.\ref{f:BH} is shown in Fig.\ref{f:WH} on right. It also shares exactly the same problem of Fig.\ref{f:BH} on right.

Whenever a BH can form and evaporate, or a WH can form and erupt, their horizon must be hanging in the middle of the Penrose diagram, and there is no way to let the event on horizon maps to the infinite Schwarzschild time.

\begin{figure}[htb]
\centering
\includegraphics[scale=0.30, angle=0]{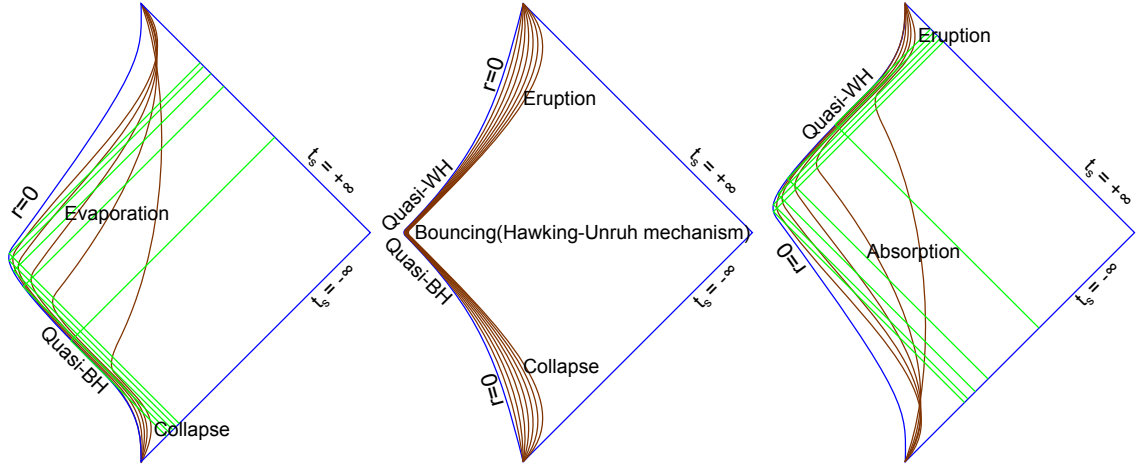}
\caption{Quasi-BH and quasi-WH} \label{f:qBHqWH}
\end{figure}

The only feasible solution is BH(WH) can never form at all, no horizon with slop $\pm 1$ except the future and past null infinity can appear in Penrose diagram. The only physically permitted possibility is the quasi-BH(quasi-WH), as shown in Fig.\ref{f:qBHqWH}.

In Fig.\ref{f:qBHqWH}, the left one shows a quasi-BH, the right one shows a quasi-WH, the middle one shows a collapsing quasi-BH bounced by some Hawking-Unruh-like mechanism and turned to an exploding quasi-WH. Because there is no essential difference between quasi-BH and quasi-WH, both of them can be called quasi-hole.

It is noteworthy that according the calculation in Sec.\ref{ss:lightcone}, the back-reaction of outgoing radiation is not necessary for preventing the formation of a BH, because the light-cone will forbid anything to enter the horizon even there is no back-reaction. However, the back-reaction of outgoing radiation should be one of the key mechanisms for the formation process and other properties of a quasi-hole. Quasi-holes may have some observable effects that distinct from perfect BH(WH).\cite{Vachaspati2007}\cite{Vachaspati2007a} It is reasonable to believe that a quasi-hole may have much more complex behaviors than a perfect BH(WH), such as bounce or burst, for a quasi-BH may have much more complex structure than a perfect BH(WH).

	\section{Summary} \label{s:summary}

In this paper, an exact universal solution with spherical symmetry of EFE was derived first, where both the Schwarzschild metric and RN metric are its special cases.

With this metric form, for a pre-existing spherical symmetric BH with finite evaporation lifespan (measured by external observers), the calculation shows nothing outside can encounter the horizon before the BH vanished, regardless of the choice of coordinates. Along same thread, we proposed a conjecture that for any pre-existing BH with a finite evaporation lifespan, no infalling object can encounter the horizon before the BH vanished, regardless of the choice of coordinates. Therefore, the widely used Penrose diagram of an evaporating BH (see Fig.\ref{f:BH}) is incorrect, because it requires that any event on horizon must mapped to some finite Schwarzschild time or Rindler time.

As a logical consequence, no perfect BH can form from gravitational collapse, regardless of the choice of coordinates. Therefore, the conundrums such as singularity and information loss are naturally avoided. The conclusion does not depend on any specific mechanism or process of BH evaporation in detail. This may imply that the conflict between quantum theory and classical gravity theory may be not as serious as it seemed to be.

However, the arguments in this paper cannot be considered as a proof because the Einstein equations need not hold when quantum effects are included, and some of the conundrums may not remain sharply in a complete unified quantum theory of gravity.

	\section{Acknowledgements}

Sincerely thanks to the referee, without his professional comments, the weakness of omitting the contribution of pressure would not be corrected. Also thanks to his constructive comments on the language clarification.

Sincerely thanks to Sheng Li of SJTU for his selfless share of inspirations, his patience of listening the stupid primitive ideas and his enthusiastic encouragements, without his illumination, the paper will not exist. Special thanks to Wei Gu, Xiaodong Li, Rongxin Miao of USTC for the help and discussion on the techniques of GR and Mathematica.

	\bibliography{ref/!}

\end{document}